# Angular Dependence of Exchange Anisotropy on Cooling Field in FM/Fluoride Thin Films

Justin Olamit,[1] Zhi-Pan Li,[2] Ivan K. Schuller,[2] and Kai Liu[1,*]

[1]*Department of Physics, University of California, Davis, CA 95616*

[2]*Department of Physics, University of California - San Diego, La Jolla, CA 92093*

(12/12/05)

**Abstract**

Exchange anisotropy in ferromagnet/antiferromagnet (FM/AF) films is usually introduced along the cooling field or FM magnetization direction. Here we investigate the dependence of the exchange anisotropy, loop bifurcation, and reversal mechanism on the cooling field direction using vector magnetometry. Three types of samples (FM=Fe, Ni /AF = $FeF_2$, $MnF_2$) have been studied where the AF layer is epitaxial (110), twinned (110), and polycrystalline. With an epitaxial AF which has one spin axis, the cooling field orients the exchange field along the spin axis. Applying the cooling field perpendicular to the spin axis results in bifurcated loops, whose shape evolves with the cooling field geometry and strength. With a twinned AF where there are two orthogonal spin axes, the exchange field direction is along the bisector of the spin axes that encompass the cooling field. With a polycrystalline AF, the exchange field direction is the same as the cooling field. Transverse hysteresis loops show that when the exchange field has a component perpendicular to the applied field, the magnetization reversal occurs by rotation in the direction of the perpendicular component. Our results demonstrate that in fluoride films, the exchange field is established primarily by the AF anisotropy direction, and only to a lesser extent the cooling field or the magnetization direction. The bifurcated loops are due to a distribution of AF anisotropies and large AF domain sizes. Furthermore, the magnetization reversal process is extremely sensitive to the exchange field direction.

## I. Introduction

There has been intense recent interest in ferromagnet / antiferromagnet (FM/AF) exchange biased systems. Exchange bias is established as the FM is cooled below the AF Néel temperature, usually in a cooling field. The exchange biased samples display a shifted hysteresis loop, by an amount known as the exchange field $H_E$, and have an enhanced coercivity $H_C$.[1,2] Angular dependence of $H_E$ and $H_C$ have been investigated both experimentally and theoretically,[3-7] revealing the unidirectional and uniaxial symmetries of $H_E$ and $H_C$, respectively. In typical angular dependence studies, the direction of the cooling field $H_{FC}$ used to bias the sample is kept constant – often along a sample anisotropy direction.

Shi and Lederman observed an extreme sensitivity of the exchange bias to the cooling field direction in thin films with a $Fe_xZn_{1-x}F_2$ AF layer.[6] They found that in samples with a twinned AF, changing the cooling field direction to an angle near one of the twin anisotropy directions caused a 90° (spin-flop-like) change in the direction of the exchange field. They attributed this to a reversal in the anisotropy direction of one of the twinned structures. In samples with an epitaxial AF layer, they found an evenly bifurcated loop (half negatively biased and half positively biased) when they field cooled 90° away from the spin axis and a 75% positively biased loop when field cooling at 91°.

Arenholz et al.[8] and Tillmanns et al.[9] have observed that in twinned Fe/MnF$_2$ films, the shape of a kinked hysteresis loop evolves sensitively with the angle between the measurement and bias fields. Other studies have examined the dependence of the magnetization reversal mechanism on the exchange field and anisotropy fields in the samples.[7, 10-13] In particular, Beckman et al.[10] and Liu et al.[7] showed that when the unidirectional exchange anisotropy is larger than the uniaxial AF anisotropy, the reversal occurs as a same-side rotation.

In this work, we investigate the dependence of exchange bias, loop bifurcation, and reversal mechanism on the angular position of the cooling field with respect to the AF anisotropy. Three types of samples were chosen because of their known bulk anisotropy directions: untwinned epitaxial AF with a uniaxial anisotropy, twinned AF with a four-fold anisotropy, and polycrystalline AF with random anisotropy. We found explicitly that the net exchange field is established *jointly* by the cooling field and anisotropy directions. In the epitaxial sample, the exchange field is along the uniaxial AF anisotropy direction favored by the cooling field. When the cooling field is applied nearly perpendicular to the spin axis, bifurcated loops are observed due to a distribution of AF anisotropy strengths and large AF domain sizes. In samples with twinned AF, the net exchange field points along the bisector of the orthogonal twin anisotropy directions which encompass the cooling field. As the cooling field direction is varied from 0° to 360°, four discrete exchange field directions are created; the exchange field changes to a new direction when the cooling field is changed to lie between a new pair of anisotropy axes. In the polycrystalline sample, the lack of an intrinsic macroscopic anisotropy direction causes the direction of the exchange field to be the same as the cooling field.

Furthermore, for all samples, transverse hysteresis loops show that when the exchange field has a component perpendicular to the applied field, the magnetization reversals occur by rotations in the direction of the perpendicular component.

## II. Experiments

Three types of samples have been studied in this work, with untwinned epitaxial AF (e-AF), twinned AF (t-AF), and polycrystalline AF (p-AF). The FM (Ni, Fe) layer is always polycrystalline.

One set of thin films of Al (76 Å) / Ni (210 Å) / FeF$_2$ (500 Å) has been grown onto single crystal MgF$_2$ (110), MgO (100) and Si (100) substrates, to respectively achieve untwinned epitaxial (110), twinned (110), and polycrystalline FeF$_2$.[6, 11, 14, 15] The Al layer serves as a capping layer. All samples have been grown by electron beam evaporation, using conditions similar to those reported in earlier publications.[11, 14-17] The FeF$_2$ layer was deposited at 300 ºC onto MgF$_2$ and at 200 ºC onto MgO and Si substrates. The Ni and Al layers were grown at 150 ºC. The crystal structures of FeF$_2$ and Ni (always polycrystalline) have been confirmed by x-ray diffraction. The FeF$_2$ (110) rocking curve full-width at half maximum (FWHM) is about 3.9º (using Cu K$\alpha$) for twinned FeF$_2$ and < 1.2º for untwinned epitaxial FeF$_2$.

Using similar conditions, we also prepared a twinned AF sample of Al (50 Å) / Fe (120Å) / MnF$_2$ (500 Å) / ZnF$_2$ (250 Å) on MgO (100) and a polycrystalline AF sample of Al (50 Å) / Fe (150 Å) / FeF$_2$ (0-200Å) on Si. The MnF$_2$ has almost identical crystal and spin structure as FeF$_2$, but different anisotropy strength.[18, 19] Both FeF$_2$ and MnF$_2$ have their spin axis along the [001] direction.

Thus for the e-AF sample (Ni/FeF$_2$), there is one in-plane spin axis; for the t-AF samples (Ni/FeF$_2$ & Fe/MnF$_2$), there are two orthogonal in-plane spin axes, 45º relative to the MgO [001] direction; and for the p-AF samples (Ni/FeF$_2$ & Fe/FeF$_2$), the spin axes are random. Table I summarizes the samples, relative film orientations on substrates, and the measurement geometries.

Field cooling and magnetic measurements were performed in a Princeton Measurement Corp. Vibrating Sample Magnetometer (VSM), equipped with a cryostat and vector detection coils. The applied field $H$ always remained in the film plane as the samples were rotated about the film normal. The epitaxial sample was mounted with the AF spin axis (MgF$_2$/FeF$_2$ [001] direction) parallel to $H$. The twinned samples were mounted so that one of the bisectors of the orthogonal twin AF spin axes (// MgO [001]) was parallel to $H$. The polycrystalline samples were mounted so that one edge of the square-shaped samples was parallel to $H$. For each sample, this starting position was designated as the 0° angular position (Table 1). The samples were held at $T > 100$ K (above the $T_{Néel}$ = 78 K of FeF$_2$ and 67 K of MnF$_2$) and rotated to a particular cooling field angle $\alpha$. A cooling field of $H_{FC}$ = 2 kOe, 1 kOe, and 2 kOe was then applied on the epitaxial, twinned and polycrystalline samples, respectively, during the cooling to 15 K [Fig. 1(a)]. The cooling field was large enough to saturate the sample, but small enough not to induce positive bias[14] or loop bifurcation.[15, 20, 21] At 15 K, while the angle was kept at $\alpha$, both longitudinal moment (the component parallel to $H$) and transverse moment (the component perpendicular to $H$) were measured with vector detection coils [Fig. 1(b)]. The sample was then rotated to the 0° (e.g., for the epitaxial sample, FeF$_2$ [001] // $H$) and 90° geometries, and again both longitudinal and transverse loops were measured [Figs.

1(c) & 1(d)]. In the following, we will denote $\theta$ as the measurement angle ($\theta = 0°$, $\alpha$, and 90°). The sample was then heated to $T > 100$ K and rotated to a new angle $\alpha$, and the field cooling and measurements were repeated. Thus each measurement is defined by both the cooling field angle $\alpha$ and the measurement angle $\theta$ (0°, $\alpha$, and 90°). On the epitaxial sample, more detailed angular dependence measurements – where $\theta$ is varied gradually after a field cool along $\alpha = 0°$ – were also measured with vector coils.

### III. Results

#### A. Epitaxial Sample

1. Angular dependence

For the Ni/epitaxial-FeF$_2$ sample, representative longitudinal and transverse magnetic hysteresis loops measured are shown in Fig. 2(a) for the $\theta = 0°$ measurement geometry when the cooling field angle $\alpha = 30°$. A large loop shift, or exchange field $H_E$, is found. Variations of $H_E$ with the cooling field angle $\alpha$ and measurement angle $\theta$ are shown in Fig. 3. When $H$ is applied along $\theta = 0°$, $H_E$ has two discrete values: about -1000 Oe for $|\alpha| < 90°$ and +1000 Oe for $|\alpha| > 90°$, with a periodicity of 360° [Fig. 3(a), blue squares]. When $H$ is applied along $\theta = 90°$, $H_E$ remains essentially zero, for all $\alpha$ [Fig. 3(a), black circles]. When $H$ is applied along $\theta = \alpha$, $H_E$ vanishes around $\alpha = \pm 90°$, with a periodicity of 180° [Fig. 3(a), red triangles].

These results suggest that the uniaxial AF spin axis and the cooling field $H_{FC}$ *jointly* dictate the exchange anisotropy. $H_{FC}$ selects one of the two opposite AF spin directions [#1 and #2 in Figs. 3(b) and 3(c)] to establish the unidirectional exchange anisotropy. For $|\alpha| < 90°$, $H_{FC}$ projects onto spin direction #1 [Fig. 3(b)] and forces $H_E$ onto it as well. As a result, $H_E$ has a constant value of about -1000 Oe when measured along $\theta = 0°$ and essentially zero when $\theta = 90°$. For $|\alpha| > 90°$, $H_{FC}$ projects, and forces $H_E$, onto spin direction #2 [Fig. 3(c)]. The opposite exchange anisotropy leads to a positive bias measured at 0°. Note that this positive bias, due to the measurement geometry, is different from that reported earlier in Fe/FeF$_2$ induced by a large $H_{FC}$.[14, 22] Along $\theta = \alpha$, $H_E$ varies in a sinusoidal manner, but its magnitude is less than or equal to the magnitude of $H_E$ measured along $\theta = 0°$. This supports the notion that the maximum exchange field is along $\theta = 0°$. The small but nonzero $H_E$ measured at $\theta = 90°$ [Fig. 3(a), black circles] indicates small misalignments of the AF spin axis with $H$.

2. Loop bifurcation

Longitudinal loops measured with $H$ applied along $\theta = 0°$ when the cooling field $H_{FC} = 2$ kOe is applied along $88° < \alpha < 92°$ are shown in Fig. 4(a). At these field cooling angles, two subloops appear, oppositely biased by the same amount, similar to previous observations.[6] When $\alpha = 88.5°$, the sample is completely negatively biased. As $\alpha$ increases, a positively biased subloop grows at the expense of the negatively biased loop. When $\alpha = 92.0°$, the sample is completely biased in the positive direction. The appearance of the subloops can be explained by the existence of large AF domains, which cause the FM layer to split into domains that are oppositely biased.[15, 20, 21] This is similar to zero-field cooled behavior previously observed, where a match of FM and AF domain sizes was important.[23-25] The large AF domains are indeed expected in the Ni/epitaxial-FeF$_2$ samples, which have been shown to exhibit a similar loop bifurcation as a function of $H_{FC}$ strength.[15] They are also consistent with loop bifurcations seen in synthetic antiferromagnets with large domain sizes.[26, 27]

The evolution of the subloop sizes with field cooling angle $\alpha$ is due to a distribution of anisotropy strengths of the AF spins. Fig. 4(b) shows longitudinal loops measured at $\theta = 0°$ when cooling field $H_{FC}$ of different strength is applied along $\alpha = 89°$. The relative sizes of the positively and negatively biased subloops change with $H_{FC}$. As $H_{FC}$ is increased, the sample becomes more positively biased. This occurs because the increasingly stronger $H_{FC}$ overcomes the antiparallel coupling between the AF and FM spins,[14] forcing the AF spins to point along the spin axis parallel to (or making a sharp angle with respect to) the applied field and leading to positive bias. In the ideal case with a uniform exchange anisotropy strength, the transition from negative to positive bias happens at a single $H_{FC}$ when Zeeman energy overcomes the interfacial exchange energy;[28] in realistic cases with a distribution of anisotropy strengths, it occurs over a range of $H_{FC}$, as shown in Fig. 4(b). When $H_{FC}$ strength is fixed and only its angle $\alpha$ changed, as depicted in Fig. 4(a), both the exchange anisotropy distribution and the geometry come into play. The latter determines an "effective" cooling field, or the projection of $H_{FC}$ onto the spin axis. Thus the portion of the sample that is positively biased, and in turn the subloop sizes, change with $\alpha$ accordingly. In short, the observation of the loop bifurcation and its evolution with geometry is a consequence of both large AF domain sizes and a distribution of anisotropy strengths.

3. Transverse loop

As shown in Fig. 2(a), the transverse moment has clear peaks during a field cycle, indicating certain amount of magnetization reversal is by rotation. For a particular measurement geometry, the direction of rotation (upward or downward) is dictated by the transverse component of $H_E$ (positive or negative), resulting in a positive or negative peak in the transverse loop. The very small coercivity implies that $H_E$ is larger than the anisotropy field, leading to same side rotations as predicted by Beckman *et al.*[10] and Liu *et al.*[7]

To probe the sensitivity of the reversal mechanism to sample orientation, more loops were measured with $\alpha = 0°$ and $H_{FC} = 1$ kOe, but the measurement angle is varied near $\theta = 0°$. The corresponding transverse loops are shown in Fig. 5. At $\theta = 1.8°$, the transverse loop shows no peaks. The exchange field is parallel to the applied field, so the spins in the FM have no preference for rotating up or down. This occurring at $\theta = 1.8°$ rather than $0°$ indicates that the sample was misaligned during mounting. The same origin is responsible for the transverse peaks seen in Fig. 2(a). With a small misalignment of the sample, the exchange field gains a transverse component which, in turn, causes the FM spins to rotate in that particular direction during magnetization reversal and the transverse loops develop large peaks. The sample fully rotates when $\theta = 6.0°$, or an additional misalignment of ~ 4°. A small misalignment of the spin axis with the applied field may account for the rotation mechanism seen by Fitzsimmons *et al.* and others in their studies of similar samples.[11, 15]

## B. Twinned Sample

For the Ni/twinned-FeF$_2$ sample, a representative magnetic hysteresis loop is shown in Fig. 2(b) when $H$ is applied along $\theta = 0°$ after field cooling along $\alpha = 35°$. Variations of $H_E$ with the cooling field angle $\alpha$ and measurement geometries are shown in Fig. 6(a). When $H$ is applied along $\theta = 0°$ (blue squares), $H_E$ shows three different trends: $H_E \sim -600$ Oe for $|\alpha| < 45°$, nearly zero for $45° < |\alpha| < 135°$, and $\sim +600$ Oe for $|\alpha| > 135°$. When $H$ is applied along $\theta = 90°$ (black circles), $H_E$ has the same three trends as in the $\theta = 0°$ geometry, but at different angles: nearly zero for $|\alpha| < 45°$ and $|\alpha| > 135°$, $\sim -600$ Oe for $45° < \alpha < 135°$ and $\sim +600$ Oe for $-135° < \alpha < -45°$. The Fe/twinned-MnF$_2$ sample exhibits similar patterns (not shown).

Common to both $\theta = 0°$ and $90°$ geometries are the four discrete plateaus (at $H_E \sim \pm 600$ Oe, separated by plateaus at $H_E \sim 0$). The exchange field value jumps every $90°$ change in $H_{FC}$ direction. This pattern is due to the four-fold symmetry of the twinned-AF spin directions, rather than the two-fold symmetry in the epitaxial AF (two plateaus with $H_E$ changing every $180°$). Furthermore, when measuring at $\theta = 0°$ results in a maximum exchange field, the measurement at $\theta = 90°$ results in an exchange field close to zero, and *vice versa*. Hence, the twinned sample can be modeled as two perpendicularly oriented epitaxial samples.

At small field cool angles ($|\alpha| < 45°$), $H_{FC}$ projects onto both directions #1 and #2 of the twinned spin structure [Fig. 6(b)]. The net $H_E$ vector is directed along the bisector of those two directions. As a result, the largest exchange field is measured at $\theta = 0°$ and no exchange field is found at $\theta = 90°$. For $45° < \alpha < 135°$, direction #2 becomes opposite to the cooling field and no longer determines an exchange field direction for its twin structure. Instead, direction #3 (antiparallel to #2) and #1 now decide the direction of the net $H_E$, whose bisector is at 90° [Fig. 6(c)]. Therefore the measurement at $\theta = 90°$ results in the largest $H_E$ and no exchange field is found at $\theta = 0°$. This change of direction of the exchange field vector from 0° to 90° is the spin-flop transition described by Shi and Lederman.[6] Similarly, when $\alpha > 135°$, directions #3 and #4 determine the net $H_E$. When measured at $\theta = 0°$, the bisector of #3 and #4 points opposite to the bisector of #1 and #2, so the maximum $H_E$ is along $\theta = 0°$ with positive values, while $H_E$ at $\theta = 90°$ is zero. The trend continues for all $\alpha$ - the bisector of the two directions that encompass the $H_{FC}$ direction becomes the exchange bias direction.

### C. Polycrystalline Sample

For the Ni/polycrystalline-FeF$_2$ sample, a representative longitudinal magnetic hysteresis loop is shown in Fig. 2(c) when $H$ is applied along $\theta = 0°$ after field cooling along $\alpha = 20°$. Variations of $H_E$ with the cooling field angle $\alpha$ and measurement angle $\theta$ are shown in Fig. 7. When $H$ is applied along the field cooling direction ($\theta = \alpha$), $H_E$ is constant at 300 Oe for all cooling field angles (red triangles). When $H$ is applied along $\theta = 0°$, $H_E$ varies with $\alpha$ as a cosine function (solid squares). When $H$ is applied along $\theta = 90°$, $H_E$ varies sinusoidally with $\alpha$. The polycrystalline Fe/FeF$_2$ exhibits similar patterns (not shown).

These results indicate that the exchange field vector remains constant along the cooling field direction in the polycrystalline samples. Therefore, when $H$ is applied along the cooling field direction ($\theta = \alpha$), the measured exchange field is always at its maximum, $H_E$; in the $\theta = 0°$ and 90° geometries, the measured exchange field is the vector projection, or $H_E \cos\alpha$ and $H_E \sin\alpha$, respectively. Note that other experiments studying the angular dependence of the exchange field have shown that the exchange field may have higher ordered cosine terms.[3, 29] For example, in NiFe/CoO, the maximum exchange field is 45° away from the cooling field direction[3].

In the transverse direction, the exchange field causes the magnetization to rotate in its direction if the exchange field is not parallel to the applied field. For loops that reverse via rotation, the reversals occur on the same side. This is consistent with the ideas of Liu *et al*. and Beckman *et al*., where same-side rotation is predicted when the exchange field is larger than the anisotropy field.[7, 10]

## IV. Discussion

The way in which the cooling field and anisotropy determine the exchange field can be understood with a simple model. The interfacial AF spins responsible for the exchange bias are influenced by the other constituents of the physical system – bulk AF spins, FM spins and cooling field. During the field cooling process, the most prominent influence on the interfacial AF spins is from the other AF spins. The large AF anisotropy[30] forces the interfacial AF spins to align with the spin axis, but they cannot discriminate between the two equivalent directions. The aligned FM moments and the cooling field break the degeneracy through their projections (however small), thus aligning the exchange field along one of the two directions. In effect, together they determine the direction of the interfacial spins.

The frozen spin direction (and resultant bias field) also explains how the magnetization reversal can be changed from domain wall motion to rotation of entire domains when the measurement angle $\theta$ is changed slightly, as shown in Fig. 5. When the exchange field is precisely parallel to the applied field, the FM moments cannot choose a rotation direction; hence they reverse via the formation of up and down domains. When the sample is rotated away from this unique geometry, the interfacial AF spins that are responsible for exchange bias now have a magnetization component perpendicular to the applied field. The resultant exchange field also has a perpendicular component that breaks the symmetry between upward and downward rotations of the FM moments, resulting in transverse hysteresis loops with large peaks, as shown in Fig. 5.

In the epitaxial film, there is only one spin axis so all interfacial AF spins must point along it. Hence, the exchange field points along the spin axis and only changes direction when the cooling field is rotated far enough to cause the interfacial spins to point along the opposite direction. The twinned samples behave like two orthogonal epitaxial samples. The resultant exchange field is the average of the two individual exchange fields; hence it lies along the bisector of the spin axes (The exchange field for the twinned sample is the average if the overlying FM domain is sensitive to all the smaller underlying twinned AF domain structures).[20, 21] As the cooling field crosses the bisector of one of the spin axes, only the exchange field along that axis flips. The average exchange field then rotates 90°, appearing as the exchange bias flop as first observed by Shi and Lederman.[6]

In polycrystalline films, there is no net spin axis because all the grains are uniformly oriented about all directions. For each grain, the cooling field and FM spins orient each interfacial AF spin along the direction of the spin axis that projects onto the cooling field. Once the direction of the interfacial spin is set, it is frozen by the other AF spins in the same grain as the AF becomes ordered. The interfacial spin acts as a constant

"external field" on the FM spins. A FM domain coupled to many AF grains is thus biased along the "average" resultant direction. As with the epitaxial sample, the exchange field can break the directional symmetry for magnetization reversals, leading to rotations when the exchange field is not parallel to the applied field during measurements.

Finally, we note that the maximum exchange field observed in the present Ni / FeF$_2$ series for the e-AF, t-AF, and p-AF sample is 1000 Oe, 600 Oe, and 300 Oe, respectively. These values are much larger than those in the Fe / FeF$_2$ series, with similar layer thicknesses, reported earlier.[11] The fact that the e-AF sample displays the largest $H_E$ is also intriguing, since it is expected to have larger AF domain sizes than the t-AF and p-AF samples.[15] Usually a larger $H_E$ is expected[31] and indeed observed[32] in systems with smaller AF domain sizes. The difference in the interfacial spin density among these samples might be a key issue worthy of further investigations.

## V. Conclusions

We showed that in exchange biased films containing fluoride AF layers, the exchange field depends on the crystallinity of the AF layer and the angle of the cooling field. The cooling field selects a symmetry direction, defined by the AF spin axes, as the exchange field direction. A film with an epitaxial AF layer had an exchange field pointing along the spin axis. Films with twinned AF structures had exchange fields pointing along the bisector of the twinned spin axes. Films with polycrystalline AF layers had exchange fields pointing along the cooling field. The cooling field changes the exchange field direction when it crosses a spin axis bisector in the epitaxial heterostructure or a spin axis of the twinned heterostructure. Bifurcated hysteresis loops were observed in the epitaxial film, whose shape is sensitive to the cooling field angle near $\alpha = 90º$, as well as the cooling field strength. The loop bifurcation was a result of both large AF domain sizes and a distribution of anisotropy strengths. Vector magnetometry revealed that the magnetization reversal was extremely sensitive to the alignment of the applied field and the exchange anisotropy, and was largely by same side rotation when the exchange field had a component perpendicular to the applied field.

## Acknowledgements

Work at UCD was supported by ACS-PRF (43637-AC10), the University of California CLE, the Alfred P. Sloan Foundation, and the NEAT–IGERT (J.O.). The acquisition of a vibrating sample magnetometer which was used extensively in this investigation was supported by NSF (EAR-0216346). Work at UCSD were supported by DOE and Cal(IT)$^2$ (Z.-P. L.). We thank Drs. E. Arenholz and D. Lederman for helpful discussions.

**Figure Captions**

FIG. 1. Sample geometries relative to the 0° line (dashed arrow, defined in Table I) during (a) field cooling along angle $\alpha$ and measurement along $\theta =$ (b) $\alpha$, (c) 0° and (d) 90°.

FIG. 2. (Color online) Longitudinal (blue squares) and transverse (red circles) magnetic hysteresis loops measured at 15 K along $\theta = 0°$ of (a) Ni/epitaxial-FeF$_2$ when $\alpha = 30°$, (b) Ni/twinned-FeF$_2$ when $\alpha = 35°$ (longitudinal only); (c) polycrystalline FeF$_2$/Fe when $\alpha = 20°$ (longitudinal only).

FIG. 3. (Color online) (a) Exchange field $H_E$ at 15 K vs. cooling field angle $\alpha$ for Ni/epitaxial-FeF$_2$, for measurements along $\theta = 0°$ (blue squares), $\theta = 90°$ (black circles), and $\theta = \alpha$ (red triangles). Dashed lines are guides to the eye. Top view schematic of the epitaxial AF sample when (b) $\alpha = 30°$ and (c) $\alpha = -120°$. The cooling field $H_{FC}$ has a projection along AF spin axis #1 (b) or #2 (c), which selects the exchange field vector (solid black arrow) to be also along this direction.

FIG. 4. (Color online) (a) Longitudinal hysteresis loops measured at $\theta = 0°$ when the field cooling angle is $88.5° \leq \alpha \leq 92.0°$. The magnetization associated with each subloop changes as $\alpha$ varies around 90°. (b) Longitudinal hysteresis loops

measured at $\theta = 0°$ when the field cooling angle is $\alpha = 89°$. As $\alpha$ is increased, the positively biased loop grows at the expense of the negatively biased loop.

FIG. 5. (Color online) Hysteresis loops measured at different $\theta$ angles when $\alpha = 0°$ and $H_{FC} = 1$ kOe. The transverse loop is flat at $\theta = 1.8°$. The transverse loop peaks grow as $\theta$ changes from 1.8° to 6.0°. A longitudinal loop measured at $\theta = 6.0°$ (black squares) is shown for comparison.

FIG. 6. (Color online) Exchange field $H_E$ at 15 K vs. cooling field angle $\alpha$ for Ni/twinned-FeF$_2$, for measurements along $\theta = 0°$ (blue squares) and 90° (black circles). Dashed lines are guides to the eye. Top view schematic of the twinned AF sample when (b) $\alpha = 20°$ and (c) $\alpha = 70°$. The cooling field $H_{FC}$ projects along directions #1 and #2 (b), or along directions #1 and #3 (c), leading to an exchange field along the 0° (b) or 90° (c) line, respectively.

Fig. 7. (Color online) Exchange field $H_E$ at 15 K vs. cooling field angle $\alpha$ for Ni/polycrystalline-FeF$_2$, for measurements along $\theta = 0°$ (solid squares), 90° (open circles), and $\alpha$ (solid triangles). Dashed lines are guides to the eye.

Table I. Summary of samples with epitaxial-AF (e-AF), twinned AF (t-AF), and polycrystalline AF (p-AF), and the mounting geometries. Dashed arrow is the 0° reference line.

| Sample | e-AF | t-AF | | p-AF |
|---|---|---|---|---|
| Geometry relative to substrate orientation (arrows along rectangle show Fe spins in FeF$_2$ and MnF$_2$) | 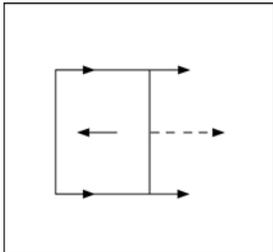 MgF$_2$/FeF$_2$ [001]→ | 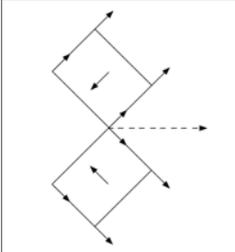 MgO [001]→ | | 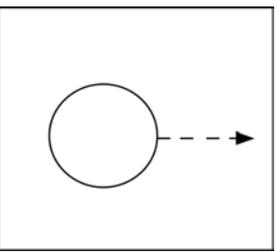 |
| AF | FeF$_2$ | FeF$_2$ | MnF$_2$ | FeF$_2$ |
| FM | Ni | Ni | Fe | Ni, Fe |
| 0º line direction | MgF$_2$/FeF$_2$ [001] | MgO [001] | | Arbitrary |

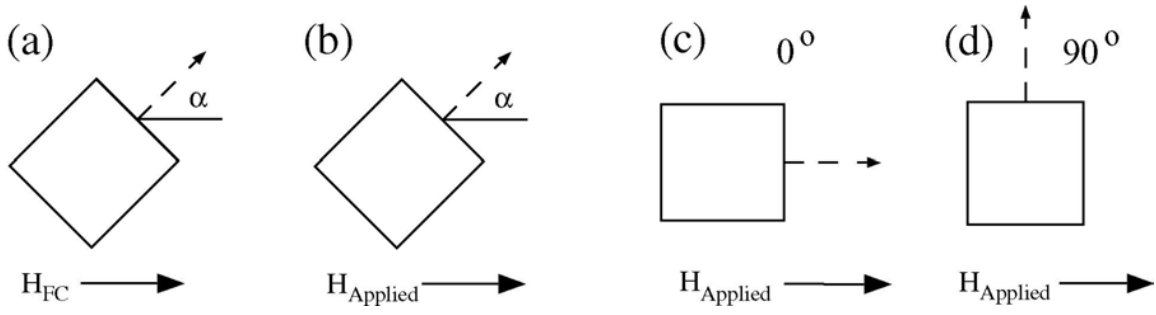

**Fig. 1**

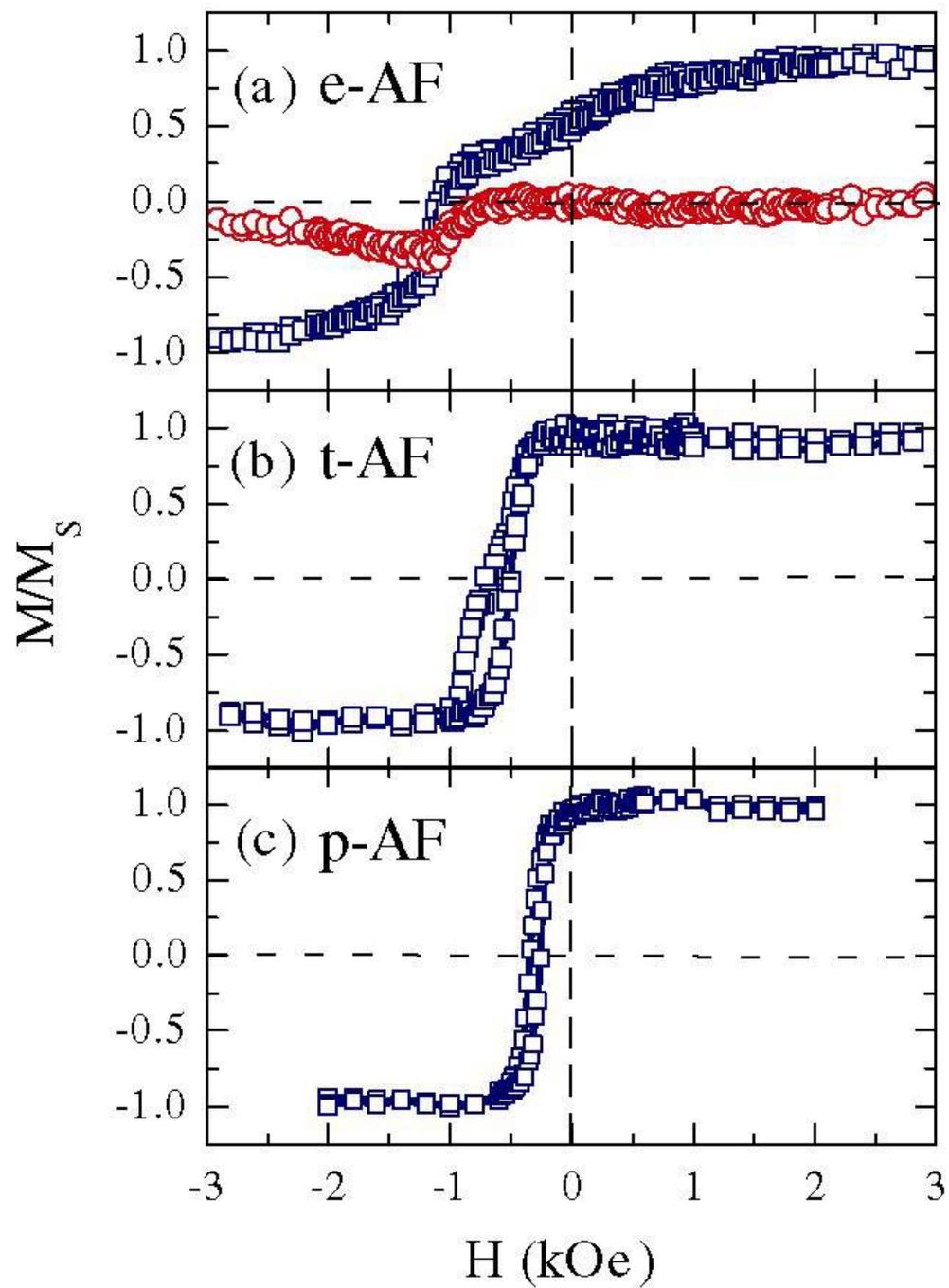

**Fig. 2**

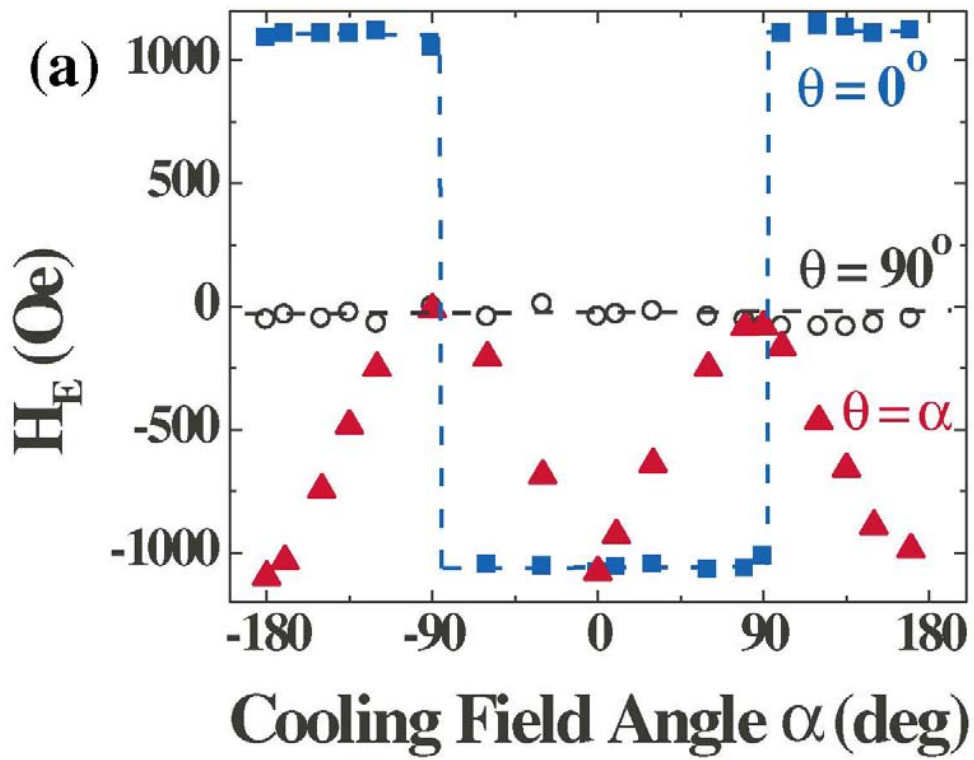

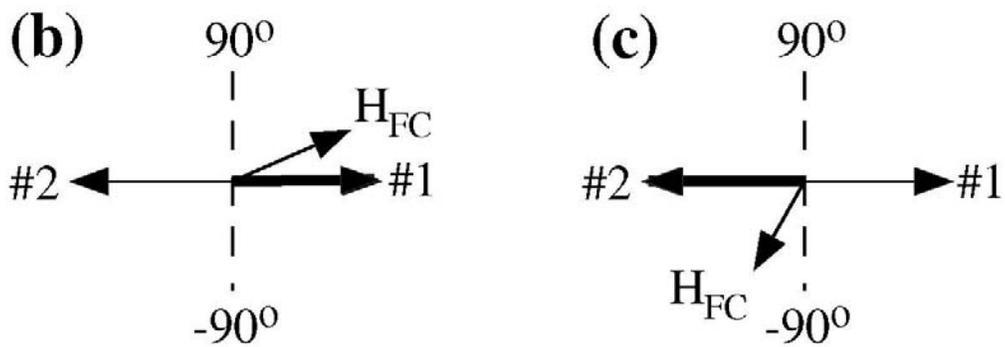

**Fig. 3**

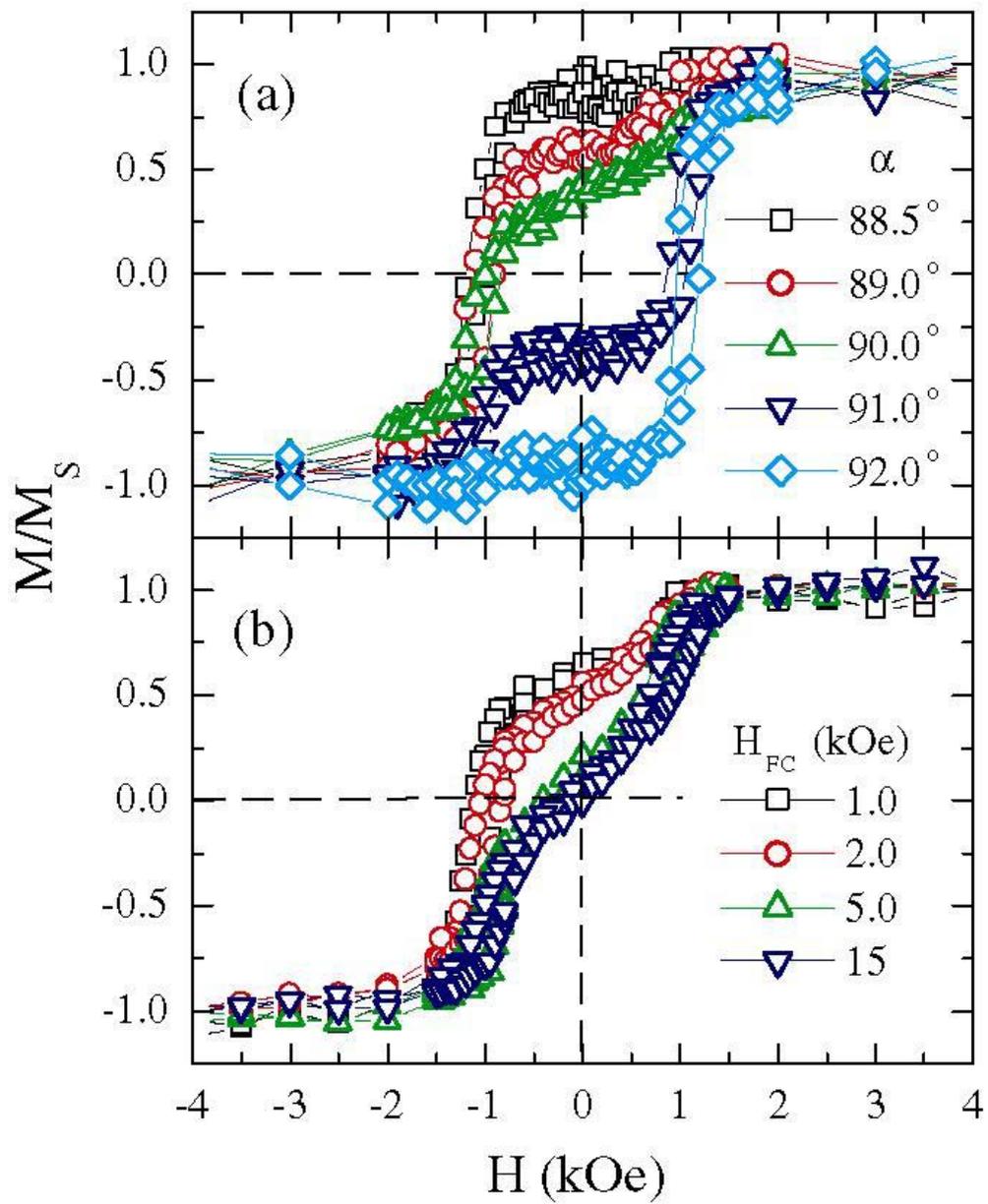

**Fig. 4**

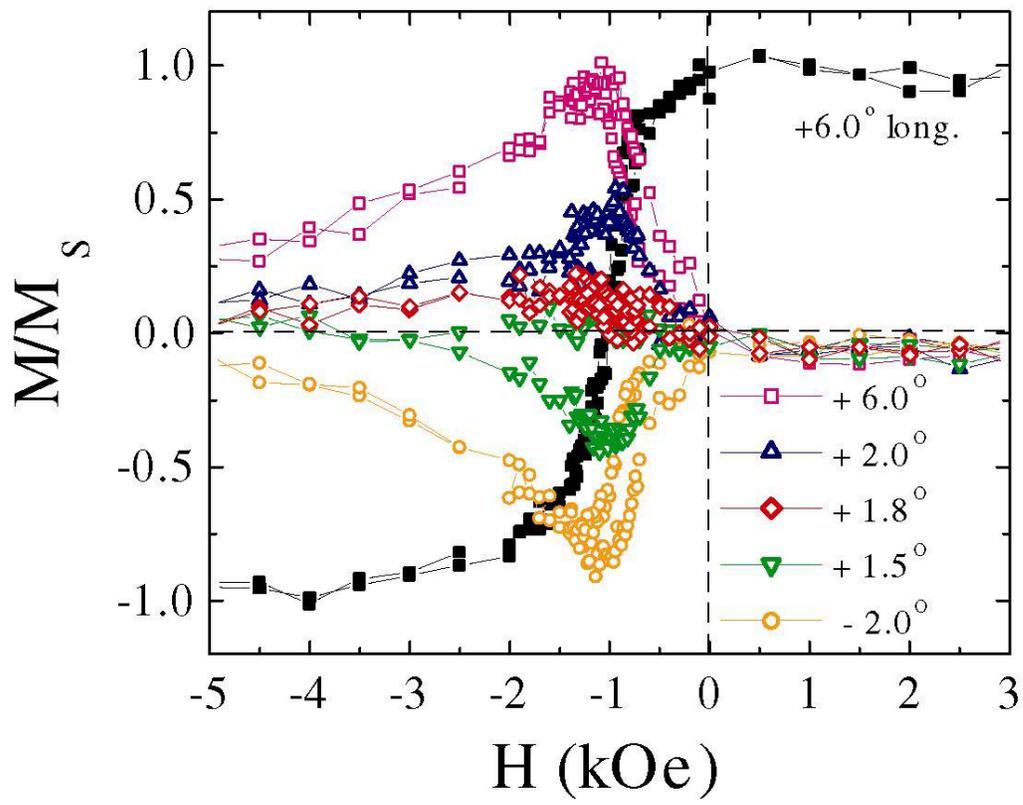

**Fig. 5**

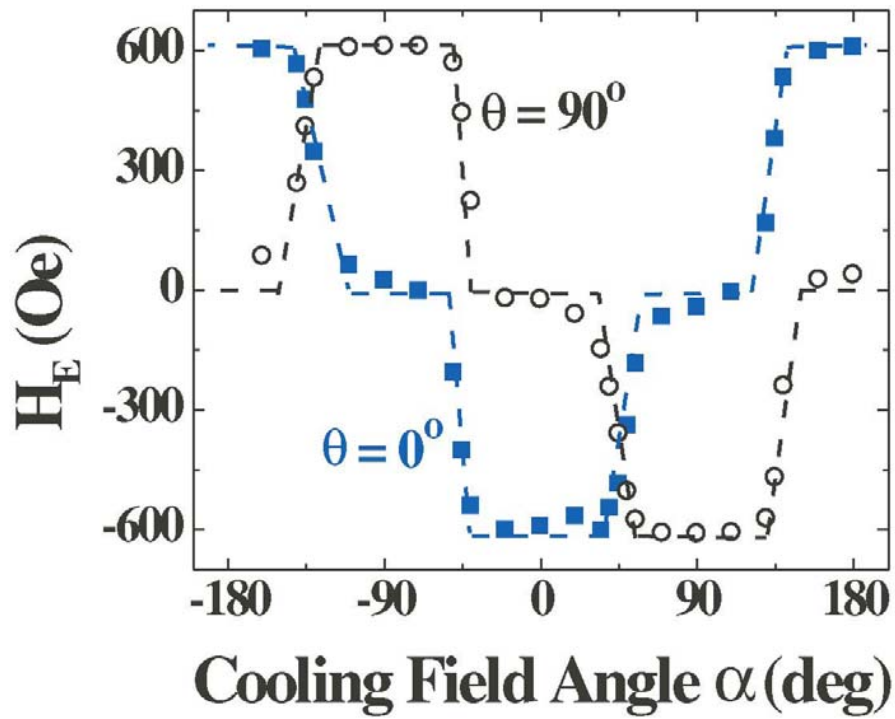
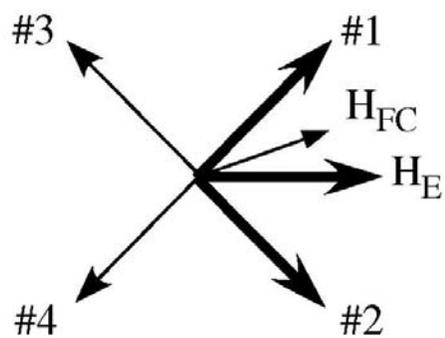
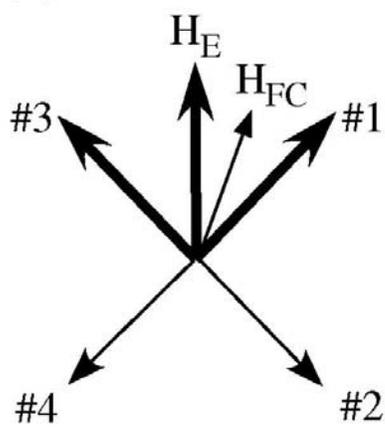

**Fig. 6**

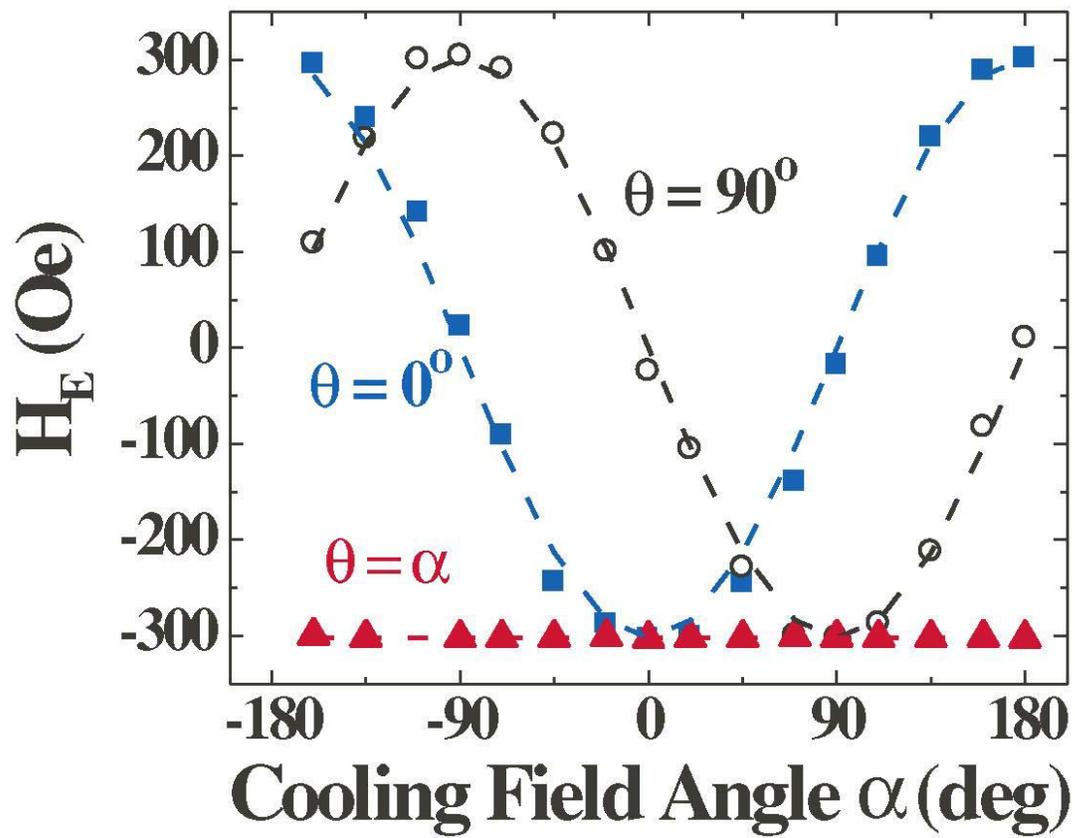

**Fig. 7**